\documentclass[twocolumn,floatfix,preprintnumbers,nofootinbib,superscriptaddress]{revtex4}

\usepackage{ulem}
\usepackage{bm}
\usepackage{times}
\usepackage{amssymb,amsbsy,amsmath,amsfonts}
\usepackage{graphicx}
\usepackage{float}
\usepackage{color}
\usepackage{morefloats}
\usepackage{rotating}
\usepackage{srcltx}
\usepackage{slashed}
\usepackage{subfigure}
\usepackage{multirow}
\usepackage{verbatim}
\usepackage{hyperref}
\usepackage{tabularx}

\usepackage[outdir=./]{epstopdf}

\begin{document}

\title{Theoretical study of the process $D^+_s \to \pi^+ K^0_S K^0_S$ and the isovector partner of $f_0(1710)$}

\author{Xin Zhu}
\affiliation{School of Physics and Microelectronics, Zhengzhou
University, Zhengzhou, Henan 450001, China} \affiliation{Institute
of Modern Physics, Chinese Academy of Sciences, Lanzhou 730000,
China}
\author{De-Min Li}\email{lidm@zzu.edu.cn}
\affiliation{School of Physics and Microelectronics, Zhengzhou
University, Zhengzhou, Henan 450001, China}
\author{En Wang}\email{wangen@zzu.edu.cn}
\affiliation{School of Physics and Microelectronics, Zhengzhou
University, Zhengzhou, Henan 450001, China}
\author{Li-Sheng Geng} \email{lisheng.geng@buaa.edu.cn}
\affiliation{School of Physics, Beihang University, Beijing 102206,
China} \affiliation{Beijing Key Laboratory of Advanced Nuclear
Materials and Physics, Beihang University, Beijing 102206, China}
\affiliation{School of Physics and Microelectronics, Zhengzhou
University, Zhengzhou, Henan 450001, China}
\author{Ju-Jun Xie} \email{xiejujun@impcas.ac.cn}
\affiliation{Institute of Modern Physics, Chinese Academy of
Sciences, Lanzhou 730000, China} \affiliation{School of Nuclear
Sciences and Technology, University of Chinese Academy of Sciences,
Beijing 101408, China} \affiliation{School of Physics and
Microelectronics, Zhengzhou University, Zhengzhou, Henan 450001,
China}

\date{\today}

\begin{abstract}

We present a theoretical study of $a_0(1710)$, the isovector partner of $f_0(1710)$, in the process $D^+_s \to \pi^+ K^0_SK^0_S$. The weak interaction
part proceeds through the charm quark decay process: $c(\bar{s}) \to (s + \bar{d} + u)(\bar{s})$, while the hadronization part takes place in two mechanisms, differing in how the quarks from the weak
decay combines into $\pi K^*$ with a quark-antiquark pair $q\bar{q}$ with the vacuum quantum
numbers. In addition to the contribution from the tree diagram of the $K^{*+} \to \pi^+ K^0_S$, we have also considered the $K^*\bar{K}^*$ final-state interactions within the chiral unitary approach to generate the intermediate state $a_0(1710)$, then it decays into the final states $K^0_SK^0_S$.  We find that the recent experimental measurements on the $K^0_SK^0_S$ and $\pi^+K^0_S$ invariant mass distributions can be well reproduced, and the proposed mechanism can provide valuable information on the nature of scalar $f_0(1710)$ and its isovector partner $a_0(1710)$.
\end{abstract}

\maketitle

\section{Introduction} \label{sec:Introduction}

Though the scalar $f_0(1710)$ resonance with $I^G(J^{PC}) = 0^+(0^{++})$ is a well established state quoted in
the Review of Particle Physics
(RPP)~\cite{ParticleDataGroup:2020ssz}, it has  attracted a lot
of discussions and debates on its structure. The main decay
channels of the $f_0(1710)$ resonance are $K\bar{K}$ and $\eta\eta$,
while the $\pi\pi$ decay branching ratio of the $f_0(1710)$
resonance is very small~\cite{ParticleDataGroup:2020ssz}. This 
indicates that $f_0(1710)$ resonance has a large $s\bar{s}$ component in its wave function.
This is indeed what was found in Ref.~\cite{Close:2005vf}. It has also been suggested as a scalar glueball candidate~\cite{Gui:2012gx,Janowski:2014ppa,Fariborz:2015dou}. Furthermore, the scalar mesons $f_0(1370)$, $f_0(1500)$, and $f_0(1710)$ cannot be simultaneously accommodated in the quark model, thus they were widely investigated by different mixing schemes~\cite{Achasov:1995nh,Amsler:1995tu,Amsler:1995td,Li:2000xy,Li:2000yn,Close:2000yk,Li:2000cj}. 

On the other hand, the $f_0(1710)$ was proposed to be a state dynamically generated from the vector meson-vector meson
interactions~\cite{Geng:2008gx,Geng:2009gb,Du:2018gyn}, which remains valid even after the vector meson-pseudoscalar meson and pseudoscalar meson-pseudoscalar meson interactions are  included in the coupled channel approach~\cite{Garcia-Recio:2013uva,Wang:2019niy,Wang:2021jub}.
Within this picture, the $f_0(1710)$ couples mostly to the 
$K^*\bar{K^*}$ channel and most properties of $f_0(1710)$ can be
well
reproduced~\cite{Wang:2021jub,Nagahiro:2008bn,Branz:2009cv,Geng:2010kma,Wang:2011tm,MartinezTorres:2012du,Xie:2014gla,Dai:2015cwa,Dai:2018thd,Molina:2019wjj}.  

In fact, in Refs.~\cite{Geng:2008gx,Du:2018gyn}, an isospin one partner of the
$f_0(1710)$ state is also obtained, with its mass around $1780$ MeV
and negative $G$-parity. The $a_0(1710)$ also couples mostly to the
$K^* \bar{K}^*$ channel, but the $\rho \omega$ and $\rho \phi$
channels are also important. Very similar conclusions are also found
in Ref.~\cite{Wang:2022pin}, where these pseudoscalar-pseudoscalar
coupled channels were taken into account, while the obtained mass of
$a_0(1710)$ of Ref.~\cite{Wang:2022pin} is smaller than those predicted in Refs.~\cite{Geng:2008gx,Du:2018gyn}. The properties of the
$a_0(1710)$ of Refs.~\cite{Geng:2008gx,Geng:2009gb,Wang:2022pin} are
collected in Table~\ref{tab:azeroparameters}, where the results of
Ref.~\cite{Wang:2022pin} are obtained with a cutoff  $q_{\rm max} =
1000$~MeV. Within the ranges of the model parameters of
Ref.~\cite{Du:2018gyn}, the $a_0(1710)$ mass is predicted in
the range of $1750 \sim 1790$ MeV. In addition, one isovector scalar resonance with a mass of 1744~MeV is also predicted within the Regge trajectories~\cite{Wang:2017pxm}.

\begin{table}[htbp]
	\begin{center}
		\caption{\label{tab:azeroparameters} Predicted properties of the
			$a_0(1710)$ state. $g_{K^*\bar{K}^*}$ stands for the coupling of $a_0(1710)$ to the $K^*\bar{K}^*$ channel. $\Gamma_{K\bar{K}}$ corresponds to the partial decay width of the $a_0(1710) \to K\bar{K}$. All are in units of MeV.}
		\begin{tabular}{lccccccccccccccccccccccccccccccccccccccccccccc}\hline\hline
			
			Set &$M_{a_0(1710)}$  & $\Gamma_{a_0(1710)}$   & $g_{K^*\bar{K}^*}$
			&$\Gamma_{K\bar{K}}$ \\ \hline
			I (Refs.~\cite{Geng:2008gx,Geng:2009gb})    & $1777$   & $148$   & $(7525,-i1529)$   & $36$           \\
			II (Ref.~\cite{Wang:2022pin})                & $1720$   & $200$   & $(8731,-i2200)$   & $74$           \\
			\hline\hline
		\end{tabular}
	\end{center}
\end{table}

Recently, the BESIII Collaboration has performed an amplitude analysis
of the process $D_s^+ \to \pi^+ K_S^0 K_S^0$~\cite{BESIII:2021anf}.
It is found that there is an enhancement in the $K^0_S
K^0_S$ invariant mass spectrum around $1.7$ GeV, which was not seen in the BESIII earlier measurements of $D_s^+\to K^+K^-\pi^+$~\cite{BESIII:2020ctr}. This indicates the existence of the isospin one partner of the $f_0(1710)$
resonance, i.e., $a_0(1710)$. In addition, the $a_0(1710)$ state was also observed in the $\pi \eta$ invariant  mass spectrum of
the $\eta_c \to \eta \pi^+\pi^-$ decay by
the {\it BABAR} Collaboration~\cite{BaBar:2021fkz}. The
Breit-Wigner mass and width of the $a_0(1710)$ state~\footnote{It should be stressed that in Ref.~\cite{BESIII:2021anf} BESIII does not distinguish between the $a_0(1710)$
and $f_0(1710)$, and denotes the combined state as
$S(1710)$.} are determined
as,
\begin{eqnarray}
M_{a_0(1710)} &=& 1723 \pm 11 \pm 2 ~{\rm MeV},\\
\Gamma_{a_0(1710)} &=& 140 \pm 14 \pm 4~{\rm MeV},
\end{eqnarray}
by BESIII~\cite{BESIII:2021anf}, and
\begin{eqnarray}
M_{a_0(1710)} &=& 1704 \pm 5 \pm 2 ~{\rm MeV},\\
\Gamma_{a_0(1710)} &=& 110 \pm 15 \pm 11~{\rm MeV},
\end{eqnarray}
by {\it BABAR}~\cite{BaBar:2021fkz}.

Based on the new measurement of BESIII~\cite{BESIII:2021anf},
Ref.~\cite{Dai:2021owu} has investigated the process $D_s^+ \to \pi^+ K_S^0 K_S^0$, where $D^+_s \to \pi^+
K^{*+}K^{*-}$, $\pi^+ K^{*0}\bar{K}^{*0}$ firstly happen,
then undergo the $K^*\bar{K}^*$ final state interaction to
give rise to the final states $K\bar{K}$. Accordingly, the $a_0(1710)$ and
$f_0(1710)$ resonances are dynamically generated from the $K^*\bar{K}^*$ final state interaction. The
production of $a_0(1710)$ and $f_0(1710)$ states in the $D_s^+ \to \pi^+ K_S^0 K_S^0$ and $D^+_s \to \pi^+ K^+ K^-$ reactions can be explained~\cite{Dai:2021owu}.

In this work, following Ref.~\cite{Dai:2021owu}, we will revisit the
process $D_s^+ \to \pi^+K_S^0 K_S^0$. In addition to the contributions
of the $a_0(1710)$ and $f_0(1710)$ states from the intermediate process $D^+_s \to \pi^+
K^*\bar{K}^*$, we will also study the contribution of $K^*$,
which could play a role in the intermediate process $D^+_s \to K^{*+} \bar{K}^0 \to \pi^+  K^0 \bar{K}^0$. We wish to go beyond the work of Ref.~\cite{Dai:2021owu} and  study the whole $K^0_S K^0_S$ and $\pi^+K^0_S$ invariant mass spectra, where we will focus on the roles played by $a_0(1710)$ and $K^{*+}$ to describe the line shapes of $K^0_S K^0_S$ and $\pi^+K^0_S$, rather than just the $f_0(1710)$ and $a_0(1710)$ contributions extracted from the experimental data, which was well described in Ref.~\cite{Dai:2021owu}.

The paper is organized as follows. In Sec. II, we present the
theoretical formalism of the $D_s^+ \to \pi^+K_S^0K_S^0$ decay, and
in Sec. III, we show our numerical results and discussions,
followed by a short summary in Sec. IV.

\section{Formalism} 

The decay $D^+_s \to \pi^+ K^0_SK^0_S$ can proceed via the $S$-wave $K^*\bar{K}^*$ final state interaction of the intermediate $D^+_s \to \pi^+ K^*\bar{K}^*$ process, or through the intermediate $K^*$ process of $D^+_s \to \bar{K}^0 K^{*+}$ with $K^{*+} \to K^0 \pi^+$ decay in $P$-wave. In the following, we will present the theoretical formalism of these two mechanisms respectively.

\subsection{The mechanism of $D^+_s \to \pi^+ K^*\bar{K}^* \to \pi^+ K^0_SK^0_S$ reaction}

\begin{figure}[htbp]
\includegraphics[scale=0.55]{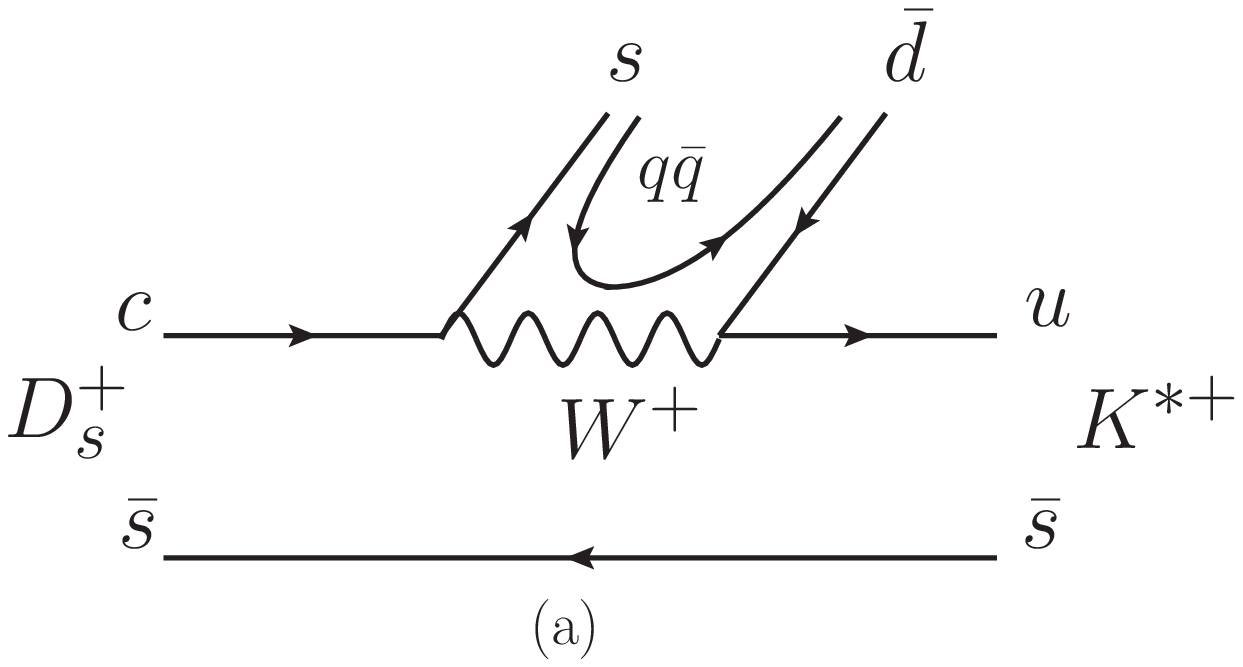}
\includegraphics[scale=0.55]{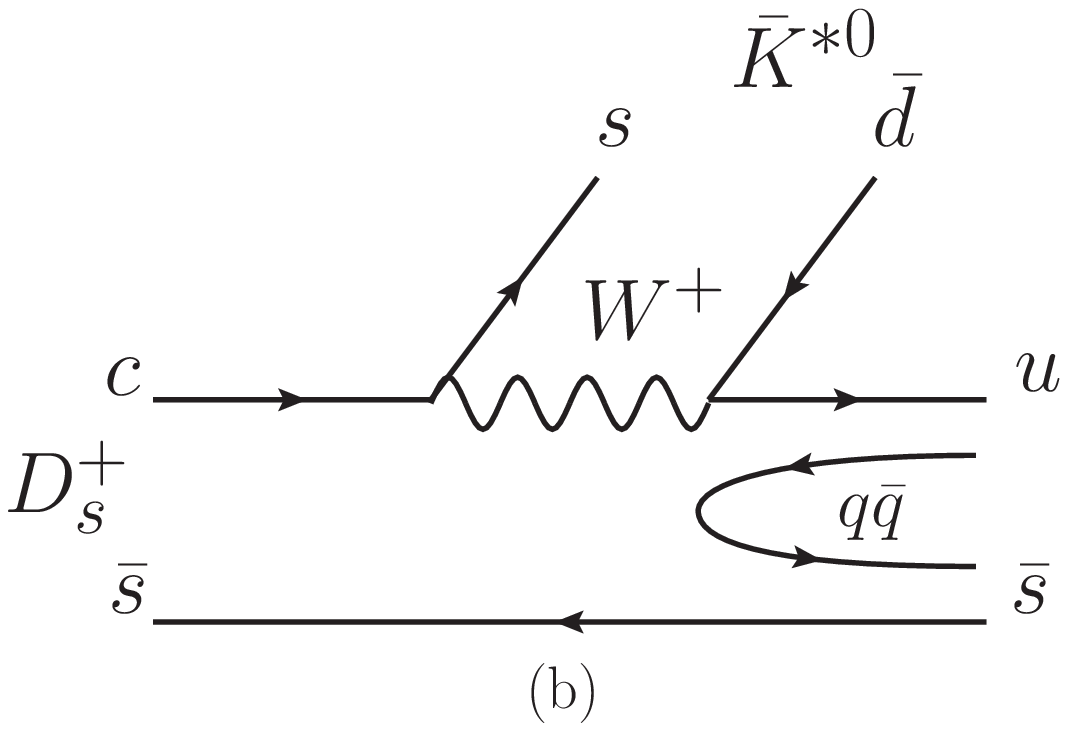}
\caption{The internal $W$ emission mechanisms for (a): $D_s^+ \to
\pi^+ K^{*-} K^{*+}$ and (b): $D^+_s \to \pi^+ K^{*0}
\bar{K}^{*0}$.} \label{Fig:Feynmanquark}
\end{figure}

As shown in Refs.~\cite{Dai:2021owu,Molina:2019udw,Wang:2021naf,Duan:2020vye}, a way for the
$D^+_s \to \pi^+ K^0_SK^0_S$ to proceed is the following: 1) the charmed quark in $D^+_s$ turns into a strange quark
with a $u\bar{d}$ pair by the weak decay shown in
Fig.~\ref{Fig:Feynmanquark}; 2) the $s \bar{d}$ (Fig.~\ref{Fig:Feynmanquark}(a)) or $u\bar{s}$ (Fig.~\ref{Fig:Feynmanquark}(b)) pair, together with the  $\bar{q}q$ ($=\bar{u}u + \bar{d}d + \bar{s}s$) pair with the vacuum quantum numbers created from vacuum, hadronizes into $(\pi K^*)^0$
or $(\pi \bar{K}^*)^+$, and the other $u\bar{s}$ and $s\bar{d}$ will hadronize to
$K^{*+}$ and $\bar{K}^{*0}$, respectively; 3) the final-state
interactions of the $K^*\bar{K}^*$ will lead to dynamical generated $a_0(1710)$, and finally it decays into $K^0_S
K^0_S$. According to the
topological classification of weak decays in
Refs.~\cite{Chau:1982da,Chau:1987tk} the above processes proceed
via the so-called internal $W$ emission mechanism.

The $D^+_s$ weak decay processes shown in Figs.~\ref{Fig:Feynmanquark} (a) and (b) can be formulated as following,
\begin{eqnarray}
D^+_s &\to& V_1 [s\bar{d} \to s(\bar{u} u + \bar{d} d + \bar{s} s)\bar{d}] (u\bar{s}\to K^{*+}), \label{eq:fig1a}\\
D^+_s &\to& V_2 [u\bar{s} \to u (\bar{u} u + \bar{d} d + \bar{s} s)\bar{s}] (\bar{d} s \to \bar{K}^{*0}), \label{eq:fig1b}
\end{eqnarray}
where $V_1$ and $V_2$ are the strength of the production vertices, and contain all the dynamical factors. One can rewrite the
two-quark two-antiquark products in the following way
\begin{eqnarray}
\sum_{i=u,d,s}{s\bar{q}_iq_i\bar{d}} &=& M_{3i}M_{i2} = (M^2)_{32}, \label{eq:sdbar} \\
\sum_{i=u,d,s}{u\bar{q}_iq_i\bar{s}} &=& M_{1i}M_{i3} = (M^2)_{13}, \label{eq:usbar}
\end{eqnarray}
where $M$ is the $q_i\bar{q}_j$ matrix in the $SU(3)$ flavor space, which is defined as
\begin{eqnarray}
	M=\left(\begin{matrix} u\bar{u} & u\bar{d} & u\bar{s}  \\
		d\bar{u}  &   d\bar{d}  &  d\bar{s} \\
		s\bar{u}  &  s\bar{d}   &    s\bar{s}
	\end{matrix}
	\right).
\end{eqnarray}

The elements of matrix $M$ can be written in terms of the pseudoscalar ($P$) or vector ($V$) mesons, which are given by~\cite{Molina:2019udw,Dai:2021owu,Wang:2021naf}.
\begin{eqnarray}
P =\left(\begin{matrix} \frac{\eta}{\sqrt{3}}+ \frac{{\pi}^0}{\sqrt{2}}+ \frac{{\eta}'}{\sqrt{6}} & \pi^+ & K^+  \\
    \pi^-  &   \frac{\eta}{\sqrt{3}}- \frac{{\pi}^0}{\sqrt{2}}+ \frac{{\eta}'}{\sqrt{6}}  &  K^0 \\
    K^-  &  \bar{K}^{0}   &    -\frac{\eta}{\sqrt{3}}+ \frac{{\sqrt{6}\eta}'}{3}
\end{matrix}
\right),
\end{eqnarray}
and
\begin{eqnarray}
V =\left(\begin{matrix} \frac{{\rho}^0}{\sqrt{2}} + \frac{{\omega}}{\sqrt{2}}  & \rho^+  & K^{*+}  \\
		\rho^-  &   - \frac{{\rho}^0}{\sqrt{2}} + \frac{{\omega}}{\sqrt{2}}  &  K^{*0} \\
		K^{*-}  &  \bar{K}^{*0}   &   \phi
	\end{matrix}
	\right).
\end{eqnarray}

The hadronization processes at the quark level in Eqs.~\eqref{eq:sdbar} and \eqref{eq:usbar} can be reexpressed at the hadronic level as,
\begin{eqnarray}
(M^2)_{32} \to (V\cdot P)_{32} &=& \pi^+K^{\ast-}-\frac{1}{\sqrt{2}}\pi^0\bar{K}^{\ast0}, \label{eq:vp32}\\
(M^2)_{13} \to (P \cdot V)_{13} &=& \pi^+K^{\ast0}+\frac{1}{\sqrt{2}}\pi^0K^{\ast+}, \label{eq:pv13}
\end{eqnarray}
where we have neglected those terms with no contribution to the intermediate $\pi K^* \bar{K}^*$ state. Using Eqs.~\eqref{eq:vp32} and \eqref{eq:pv13}, we can rewrite Eqs.~\eqref{eq:fig1a} and \eqref{eq:fig1b} as
\begin{eqnarray}
D^+_s  &\to&  V_1( \pi^+K^{\ast-}K^{\ast+}-\frac{1}{\sqrt{2}}\pi^0\bar{K}^{\ast0}K^{\ast+}),\label{eq:sdbarhadron} \\
D^+_s  & \to &  V_2 ( \pi^+K^{\ast0}\bar{K}^{\ast0}+\frac{1}{\sqrt{2}}\pi^0K^{\ast+}\bar{K}^{\ast0}). \label{eq:usbarhadron}
\end{eqnarray}

To study the decay $D^+_s \to \pi^+ a_0(1710)$ with $a_0(1710)$ dynamically generated from  the final-state interaction of $K^*\bar{K}^*$, we should sum Eqs.~\eqref{eq:sdbarhadron} and \eqref{eq:usbarhadron} and produce the combination of $K^{*+}K^{*-}$ and $K^{*0}\bar{K}^{*0}$ in isospin $I = 1$. With the isospin doublet $(K^{*+}, K^{*0})$ and $(\bar{K}^{*0}, -K^{*-})$~\cite{Close:1979bt}, we obtain,
\begin{eqnarray}
|K^{\ast0}\bar{K}^{\ast0}\rangle &=& {\frac{1}{\sqrt{2}}}(|K^\ast\bar{K}^\ast,I=1\rangle-|K^\ast\bar{K}^\ast,I=0\rangle), \nonumber \\
|K^{\ast+}K^{\ast-}\rangle &=& -{\frac{1}{\sqrt{2}}}(|K^\ast\bar{K}^\ast,I=1\rangle+|K^\ast\bar{K^\ast},I=0\rangle).  \nonumber
\end{eqnarray}

Thus, we have,
\begin{eqnarray}
\!\!\!\!\!\!\!\! & & \!\! V_1 K^{\ast+}K^{\ast-} + V_2 K^{\ast0}\bar{K}^{\ast0} \nonumber \\
\!\!\!\!\!\!\!\!	&=&  \!\!- {\frac{V_1}{\sqrt{2}}}\left(|K^\ast\bar{K^\ast},I=1\rangle+|K^\ast\bar{K^\ast},I=0\rangle \right)\nonumber \\ 
\!\!\!\!\!\!\!\!	&& \!\! + \frac{V_2}{\sqrt{2}}\left(|K^\ast\bar{K^\ast},I=1\rangle-|K^\ast\bar{K^\ast},I=0\rangle\right)\nonumber \\
\!\!\!\!\!\!\!\!	& =& \!\! {\frac{V_2-V_1}{\sqrt{2}}}|K^\ast\bar{K^\ast},I=1\rangle-{\frac{V_2+V_1}{\sqrt{2}}}|K^\ast\bar{K^\ast},I=0\rangle . \label{eq:KstarKbarstarcombination}
\end{eqnarray}
We see that the phases of the above two terms have different signs. If one term is dominant, the other one could be small and can be neglected. In this work, we will focus on the contribution from $a_0(1710)$ and ignore the $f_0(1710)$ contribution. This seems to be a reasonable choice given the reasonable description of the invariant $K^0_SK^0_S$ and $\pi^+K^0_S$ mass distributions as shown below.

\begin{figure}[tbhp]
    \centering
        \includegraphics[scale=0.49]{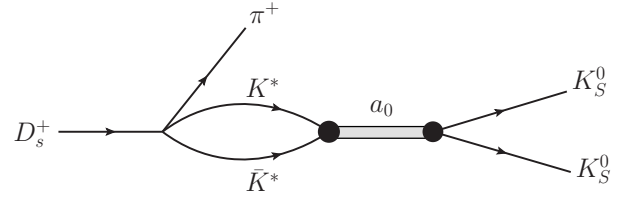}
    \caption{The diagram for the $K^*\bar{K}^*$ final state interaction for the $D^+_s \to \pi^+ K^*\bar{K}^* \to \pi^+ a_0(1710) \to \pi^+ K^0_SK^0_S$ decay.}
  \label{fig:hadronFSI}
\end{figure}

After the production of the $K^*\bar{K}^*$ pair, the final-state interaction in $S$-wave between $K^*$ and $\bar{K}^*$ takes place, in which the $a_0(1710)$ is produced, and then it decays to $K^0_SK^0_S$ in the final state.~\footnote{Note that the parameters $V_1$ and $V_2$ are assumed to be independent of the final-state interactions.} In Fig.~\ref{fig:hadronFSI} we show the rescattering diagram for the $D^+_s \to \pi^+ K^*\bar{K}^* \to \pi^+ a_0(1710) \to \pi^+ K^0_SK^0_S$ decay.

With the above formalism, the decay amplitude of the process shown in Fig.~\ref{fig:hadronFSI} can be written as~\footnote{We take $|K^0\rangle = \frac{1}{\sqrt{2}}
	(|K^0_S\rangle + |K^0_L\rangle)$ and $|\bar{K}^0\rangle =
	\frac{1}{\sqrt{2}} (|K^0_S\rangle - |K^0_L\rangle)$, where we have
	ignored the effect of $CP$ violation.}
\begin{eqnarray}
\mathcal{M}_a &=& {\frac{V_2-V_1}{4}}{\tilde{G}}_{K^{\ast}\bar{K}^{\ast}}(M_{K^0_SK^0_S}) \times \nonumber \\
&&\!\!\!\!\!\!\!\!{\frac{g_{K^\ast\bar{K^\ast}}g_{K\bar{K}}}{M_{K_S^0K_S^0}^2-M_{a_0(1710)}^2+iM_{a_0(1710)}\Gamma_{a_0(1710)}}}, \label{eq:ma}
\end{eqnarray}
where $M_{K^0_SK^0_S}$ is the invariant mass of the $K^0_SK^0_S$ system. We define $V_P = V_2 - V_1$, which will be determined from the branching fraction Br$(D^+_s \to \pi^+ K^0_S K^0_S)$. 

The $\tilde{G}_{K^*\bar{K}^*}$ is the loop function for the $K^*\bar{K}^*$ pair, which depends on $M_{K^0_SK^0_S}$. Since the $K^*$ and $\bar{K}^*$ have large total decay widths, they should be taken into account. For that purpose, the $G_{K^*\bar{K}^*}$ is not $\tilde{G}$, the loop function of two stable particles of masses $m_1$ and $m_2$, but convoluted in the masses $m_1$ and $m_2$ with the mass distributions of $K^*$ and $\bar{K}^*$ vector mesons, which can be done following Refs.~\cite{Geng:2008gx,Molina:2008jw,Xie:2013ula},
\begin{eqnarray}
	{G}_{K^{\ast}\bar{K}^{\ast}}(M_{K^0_SK^0_S}) &=& \int_{m_{-}^2}^{m_{+}^2} \int_{m_{-}^2}^{m_{+}^2}d\tilde{m_1}^2d\tilde{m}_2^2 \times \nonumber \\  
	&& \!\!\!\!\!\!\!\!\!\!\!\!\!\!\!\!\!\!\!\! \omega(\tilde{m}_1^2)\omega(\tilde{m}_2^2)\tilde{G}(M_{K^0_SK^0_S},\tilde{m}_1^2,\tilde{m}_2^2),
\end{eqnarray}
with
\begin{eqnarray}
\omega(\tilde{m}_1^2) &=& {\frac{1}{N}}\text{Im}\left(\frac{1}{\tilde{m}_1^2-m_{K^{\ast}}^2+i\Gamma(\tilde{m}_1^2)\tilde{m}_1}\right)\\
N &=& \int_{\tilde{m}_{-}^2}^{\tilde{m}_{+}^2}d\tilde{m}_1^2\text{Im}\left(\frac{1}{\tilde{m}_1^2-m_{K^{\ast}}^2+i\Gamma(\tilde{m}_1^2)\tilde{m}_1}\right),
\end{eqnarray}
and
\begin{eqnarray}
\Gamma(\tilde{m}_1^2) &=& \Gamma_{K^\ast}\frac{\tilde{k}^3}{k^3}, \\
\tilde{k} &=& \frac{\lambda(\tilde{m}_1^2,m_{\pi}^2,m_K^2)}{2\tilde{m}_1},
\end{eqnarray}
where the K${\ddot{a}}$llen function $\lambda(x,y,z) =x^2+y^2+z^2-2xy-2xz-2yz$. In this work, we take $m_+^2=\left(m_{K^\ast}+2\Gamma_{K^\ast}\right)^2$,   $m_-^2=\left(m_{K^\ast}-2\Gamma_{K^\ast}\right)^2$, $m_\pi = 138.04$~MeV and 
$m_K = 495.644$~MeV. In addition, the masses, widths and spin-parities of the involved particles are listed in Table~\ref{tab:particleparameters}.

\begin{table}[htbp]
\caption{Masses, widths and spin-parities of the involved particles in this work.}	\label{tab:particleparameters}
	\begin{tabular}{cccc}
		\hline\hline  
		Particle & Mass (MeV) & Width (MeV) & Spin-parity ($J^P$) \\ \hline
		$D^+_s$  &1968.34    &1.31 $\times 10^{-9}$             &$0^-$      \\
		$\pi^+$  &139.5704   &---             &$0^-$   \\
		$K^0_S$  &497.611     &---            &$0^-$ \\
        $K$      &495.644      &---           &$0^-$ \\
         $K^{*}$  &893.605     &49.05          &$1^-$ \\
        $K^{*+}$ &891.66      &50.8           &$1^-$ \\
        $K^{*0}$ &895.55      &47.3           &$1^-$ \\
		\hline\hline
	\end{tabular}
\end{table}

In the dimensional regularization scheme, $\tilde{G}(s=M^2_{K^0_SK^0_S},m_1^2,m_2^2)$ can be written as~\cite{Molina:2008jw, Xie:2013ula}
\begin{eqnarray}
	\tilde{G} &=& \frac{1}{16\pi^2}\Bigg\{a_{\mu} + \text{In}\frac{m_1^2}{\mu^2}+\frac{m_2^2-m_1^2+s}{2s}\text{In}\frac{m_2^2}{m_1^2} \nonumber \\
	&& \frac{p}{\sqrt{s}}\bigg[\text{In}\left(s-\left(m_2^2-m_1^2\right)+2p\sqrt{s}\right)\nonumber \\
	&& +\text{In}\left(s+\left(m_2^2-m_1^2\right)+2p\sqrt{s}\right)\nonumber \\
	&& -\text{In}\left(-s+\left(m_2^2-m_1^2\right)+2p\sqrt{s}\right)\nonumber \\
	&& -\text{In}\left(-s-\left(m_2^2-m_1^2\right)+2p\sqrt{s}\right)\bigg]\Bigg\}
\end{eqnarray}
with
\begin{eqnarray}
	p=\frac{\lambda^{1/2}(s,m_1^2,m_2^2)}{2\sqrt{s}},
\end{eqnarray}
where $\mu$ is a scale of dimensional regularization, and $a_{\mu}$ is the subtraction constant. We take $\mu = 1000$ MeV and $a_{\mu} =-1.726$ as  in Ref.~\cite{Geng:2008gx}. It is worth mentioning that the only parameter dependent part of $\tilde{G}$ is $a_\mu + {\rm ln}(m^2_1/\mu^2)$. Any change in $\mu$ is
reabsorbed by a change in $a_\mu$ through $a_{\mu'} - a_\mu = {\rm ln}(\mu'^{2}/\mu^2)$, so that the loop function $\tilde{G}$ is scale independent. It should be noted that the loop function $\tilde{G}$ can be also regularized with the cutoff method as in Refs.~\cite{Wang:2021jub,Wang:2022pin,Wang:2019niy,Lu:2020ste,Oller:1997ti,Oset:1997it}.

\begin{figure}[htbp]
	\centering
	\includegraphics[scale=0.33]{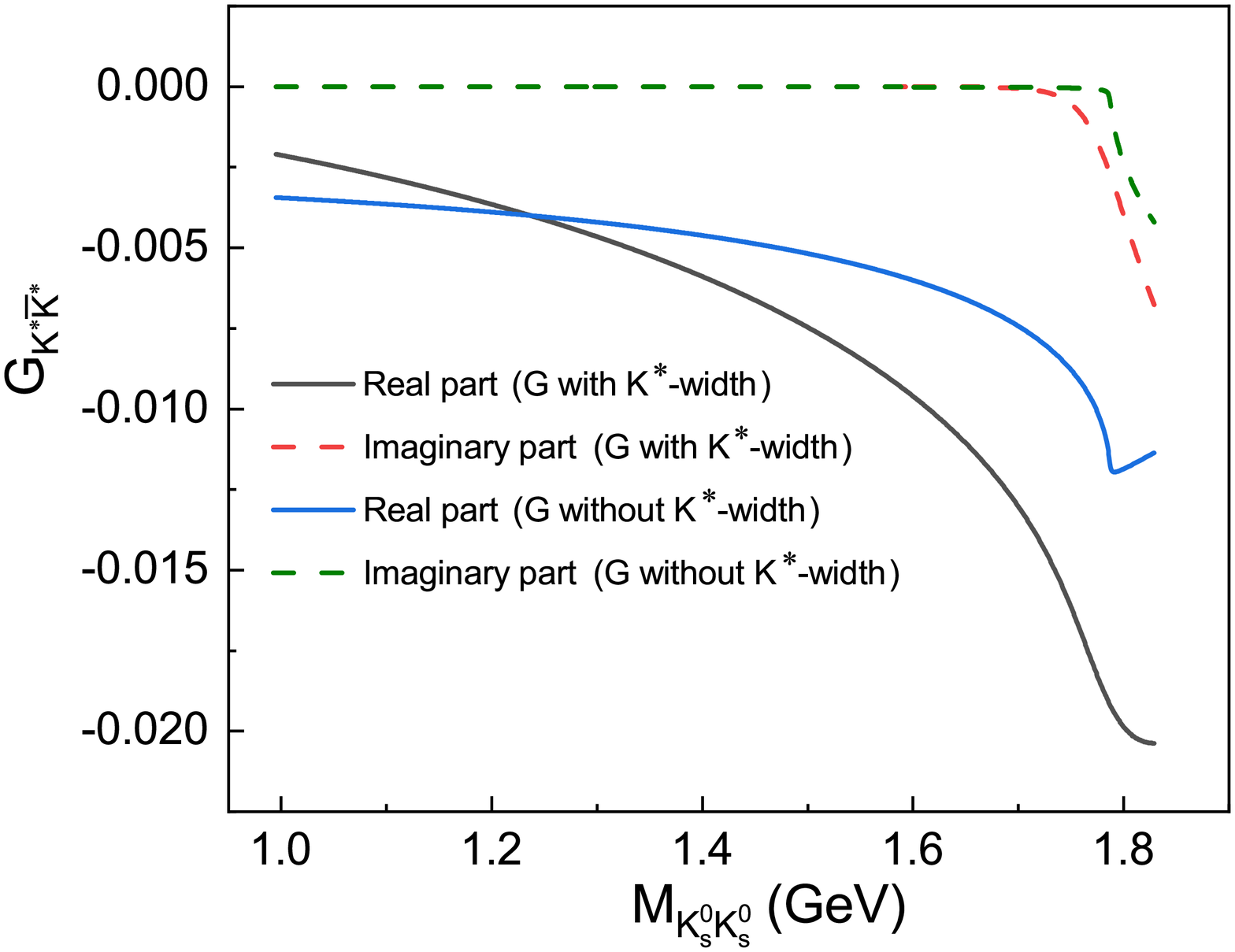}
	\vspace{-0.5cm}
	\caption{Real and imaginary parts of the loop function $G_{K^*\bar{K}^*}$ as a function of the invariant $K^0_SK^0_S$ mass computed in the dimensional regularization method and the cutoff method.}
	\label{fig:GKstarKbarstar}
\end{figure}

The so obtained real (solid curves) and imaginary (dashed curves) parts of the loop function $G_{K^*\bar{K}^*}$ as a function of the  $K^0_SK^0_S$ invariant mass are shown in Fig.~\ref{fig:GKstarKbarstar}. The results considering the $K^*$ width are obtained with the dimensional regularization method as in Ref.~\cite{Geng:2008gx}, while the results without considering the $K^*$ width are calculated with the cutoff parameter of Ref.~\cite{Wang:2022pin}.

In addition, the $g_{K\bar{K}}$ in Eq.~\eqref{eq:ma} is the coupling constant of $a_0(1710)$ to the $K\bar{K}$ channel, and it can be determined from the partial decay width of $a_0(1710) \to K\bar{K}$, which is given by
\begin{eqnarray}
\Gamma_{K\bar{K}} = \frac{g^2_{K\bar{K}}}{8\pi} \frac{p_K}{M^2_{a_0(1710)}},
\end{eqnarray}
where $p_K$ is the three momentum of the $K$ or $\bar{K}$ meson in the $a_0(1710)$ rest frame. With these $\Gamma_{K\bar{K}}$ values of Refs.~\cite{Geng:2008gx,Geng:2009gb} and Ref.~\cite{Wang:2022pin} shown in Table~\ref{tab:azeroparameters}, we obtain $g_{K\bar{K}} = 1966$ MeV and $2797$ MeV for Set I and Set II, respectively. Note that from the partial decay width, one can only obtain the absolute value of the coupling constant, but not the phase. In this work, we assume that $g_{K\bar{K}}$ is real and positive.

\subsection{The mechanism of $D^+_s \to \bar{K}^0 K^{*+} \to \pi^+ K^0_SK^0_S$ reaction}

\begin{figure}[htbp]
	\centering
	\includegraphics[scale=0.55]{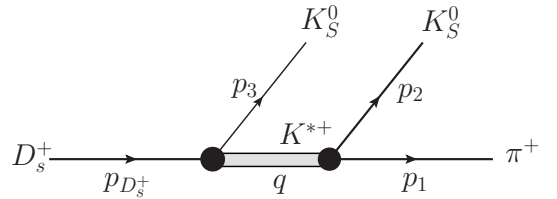}
	\caption{The decay $D_s^+\rightarrow\pi^+K_S^0K_S^0$ via the intermediate vector $K^{\ast+}$. We also show the definition of the kinematical ($p_1$, $p_2$, $p_3$, $p_{D^+_s}$) variables that we use in the present calculation.}	\label{fig:mb}
\end{figure}

In this section, we will present the formalism for the decay $D_s^+ \to \pi^+K_S^0K_S^0$ via the intermediate meson $K^{*+}$. According to the RPP~\cite{ParticleDataGroup:2020ssz}, the absolute branching fraction of the decay mode $D^+_s \to \bar{K}^0 K^{*+}$ is $(5.4 \pm 1.2)\%$, which is comparable
to the absolute branching fraction of $D^+_s \to \eta \rho^+$ that is $(8.9 \pm 0.8)\%$. As a result, the $D_s^+ \to K^0_S K^{*+}$ is important to produce $\pi^+K_S^0K_S^0$ in the final state through $K^{*+} \to \pi^+ K^0_S$ in $P$-wave, as shown in Fig.~\ref{fig:mb}.

 The decay amplitude for $D^+_s \to \pi^+ K^0_SK^0_S$ from the process shown in Fig.~\ref{fig:mb} can be obtained as
\begin{eqnarray}
    \mathcal{M}_b&=&\frac{g_{D_s\bar{K}K^*}g_{K^{\ast}K\pi}}{2}\frac{1}{q^2-m_{K^{\ast+}}^2+im_{K^{\ast+}}\Gamma_{K^{\ast+}}}\nonumber \\
    &&\times\Bigg[\left(m_{K_S^0}^2-m_{\pi^+}^2\right)\left(1-\frac{q^2}{m_{K^{\ast+}}^2}\right)\nonumber \\
    &&+2p_1\cdot p_3 \frac{m_{\pi^+}^2-m_{K_S^0}^2-m_{K^{\ast+}}^2}{m_{K^{\ast+}}^2}\nonumber \\ 
&&+2p_2 \cdot p_3 \frac{m_{\pi^+}^2-m_{K_S^0}^2+m_{K^{\ast+}}^2}{m_{K^{\ast+}}^2} \Bigg] \nonumber \\
    &&+ ({\rm exchange~ term ~with}~ p_2 \leftrightarrow p_3),
\end{eqnarray}
where $q^2 = (p_1 + p_2)^2 = M_{{\pi}K_S^0}^2$ is the invariant mass squared of the $\pi^+ K^0_S$ system. The $g_{D_s\bar{K}K^*}$ and $g_{K^*K\pi}$ denote the coupling constants of $D^+_s\to \bar{K}^0 K^{*+}$ and $K^{*+} \to K^0\pi^+$, respectively. With the masses of these particles given in Table~\ref{tab:particleparameters}, the branching fraction of Br$(D^+_s \to \bar{K}^0 K^{*+}) = (5.4 \pm 1.2)\%$ and the
partial decay width $K^{*+} \to K^0 \pi^+$ quoted in the RPP~\cite{ParticleDataGroup:2020ssz}, we obtain $g_{D_s\bar{K}K^*} = (1.05 \pm 0.12) \times 10^{-6}$ and $g_{K^* K\pi} = 3.26$. Again, we assume that $g_{D_s\bar{K}K^*}$ and $g_{K^* K\pi}$ are real and positive~\cite{Ling:2021qzl}. The uncertainty of $g_{D_s\bar{K}K^*}$ originates from the uncertainty of the branching fraction Br$(D^+_s \to \bar{K}^0 K^{*+})$, while the uncertainty of $g_{K^* K\pi}$ is ignored, since it is very small.

\subsection{Invariant mass distributions}

We can write the total decay amplitude of $D^+_s \to \pi^+ K^0_SK^0_S$ as follows, 
\begin{eqnarray}
 \mathcal{M} =\mathcal{M}_a+\mathcal{M}_b,
\end{eqnarray}
and the double differential width of the decay $D^+_s \to \pi^+ K^0_SK^0_S$ is
\begin{eqnarray}
     \frac{d^2\Gamma}{dM_{K_S^0K_S^0}{dM_{{\pi}K_S^0}}}=\frac{M_{K_S^0K_S^0}M_{{\pi}K_S^0}}{128\pi^3m_{D_s^+}^3}(|\mathcal{M}_a|^2  +|\mathcal{M}_b|^2), \label{eq:dgammadm12dm23}
\end{eqnarray}
where the interference between ${\cal M}_a$ and ${\cal M}_b$ is neglected, since these coupling constants are assumed to be real and positive, as discussed above.

In Ref.~\cite{BESIII:2021anf}, by considering the interference term between $a_0(1710)$ and $K^{*+}$, the extracted branching fraction Br$(D^+_s \to \bar{K}^0 K^{*+})$ is $(1.8 \pm 0.2 \pm 0.1)\%$, which  deviates from the CLEO result of Br$(D^+_s \to \bar{K}^0 K^{*+}) = (5.4 \pm 1.2)\%$~\cite{CLEO:1989zcg}. In this work, since the interference term is not included, we use the CLEO result to determine the value of the coupling constant $g_{D_s\bar{K}K^*}$.

Finally, one can easily obtain ${d\Gamma}/{dM_{K_S^0K_S^0}}$ and ${d\Gamma}/{dM_{{\pi}K_S^0}}$, by integrating Eq.~\eqref{eq:dgammadm12dm23} over each of the invariant mass variables with the limits of the Dalitz Plot given in the RPP~\cite{ParticleDataGroup:2020ssz}. For example, the upper and lower limits for $M_{\pi^+K^0_S}$ are as follows:
\begin{eqnarray}
\left(M_{\pi^+K_S^0}^2\right)_\text{max} &= &\left(E_{\pi^+}^\ast+E_{K_S^0}^\ast\right)^2 -  \nonumber \\
    && \left(\sqrt{E_{\pi^+}^{\ast2}-m_{\pi^+}^2}-\sqrt{E_{K_S^0}^{\ast2}-m_{K_S^0}^2}\right)^2 \nonumber \\
\left(M_{\pi^+K_S^0}^2\right)_\text{min} &=&\left(E_{\pi^+}^\ast+E_{K_S^0}^\ast\right)^2 -  \nonumber \\
    &&\left(\sqrt{E_{\pi^+}^{\ast2}-m_{\pi^+}^2}+\sqrt{E_{K_S^0}^{\ast2}-m_{K_S^0}^2}\right)^2, \nonumber
\end{eqnarray}
where the $E_{\pi^+}^\ast$ and $E_{K_S^0}^{\ast}$ are the energies of $\pi^+$ and $K_S^0$ in the $K_S^0K_S^0$ rest frame, respectively,
\begin{align}
    &E_{\pi^+}^\ast=\frac{m_{D_s^+}^2-M_{K_S^0K_S^0}^2-m_{\pi^+}^2}{2M_{K_S^0K_S^0}},  \nonumber \\
    &E_{K_S^0}^\ast=\frac{M_{K_S^0K_S^0}^2-m_{K_S^0}^2+m_{K_S^0}^2}{2M_{K_S^0K_S^0}}.
\end{align}
Similarly, one can obtain the upper and lower limits of $M_{K^0_SK^0_S}$. 

\begin{figure}[htbp]
	\centering
	\includegraphics[scale=0.32]{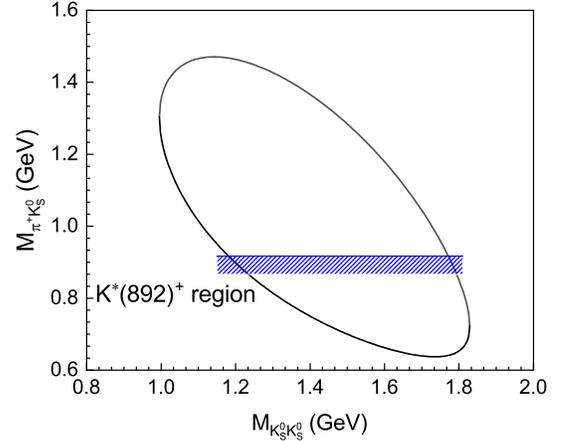}
	\vspace{-0.5cm}
	\caption{Dalitz Plot for the decay $D_s^+\rightarrow\pi^+K_S^0K_S^0$.}
	\label{fig:DalitzPlot}
\end{figure}

In Fig.~\ref{fig:DalitzPlot} we show the Dalitz Plot of the $D^+_s \to \pi^+ K^0_S K^0_S$ reaction. The blue band stands for the $K^{*+}$ region in the $\pi^+K^0_S$ channel. One can see that the $K^{*+}$ energy region overlaps largely with the $a_0(1710)$ state in the $K^0_SK^0_S$ channel.

Because the factor $V_P$ is unknown, we determine it from the branching fraction of $D^+_s \to \pi^+ K^0_SK^0_S$, which is $(0.68 \pm 0.04 \pm 0.01)\%$~\cite{BESIII:2021anf}. With the $a_0(1710)$ parameters given in Table~\ref{tab:azeroparameters}, we obtain
\begin{eqnarray}
V_P = (1.69 \pm 0.55) \times 10^{-4}.
\end{eqnarray}
for Set I, and
\begin{eqnarray}
V_P = (1.95 \pm 0.64) \times 10^{-4}.
\end{eqnarray}
for Set II.

\section{Results and Discussion} \label{sec:Results}

\begin{figure}[tbhp]
    \begin{center}
        \includegraphics[scale=0.32]{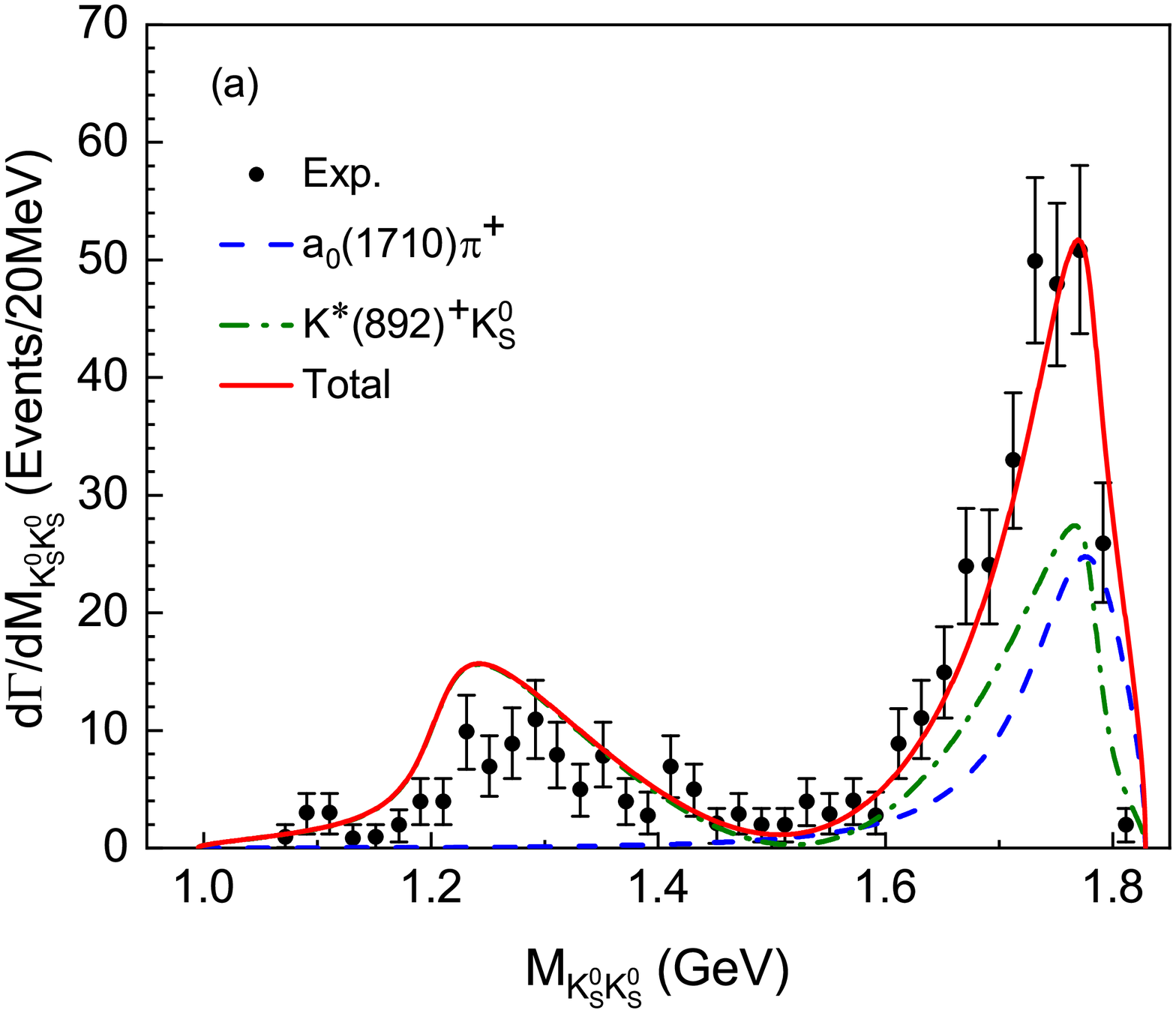}
\includegraphics[scale=0.32]{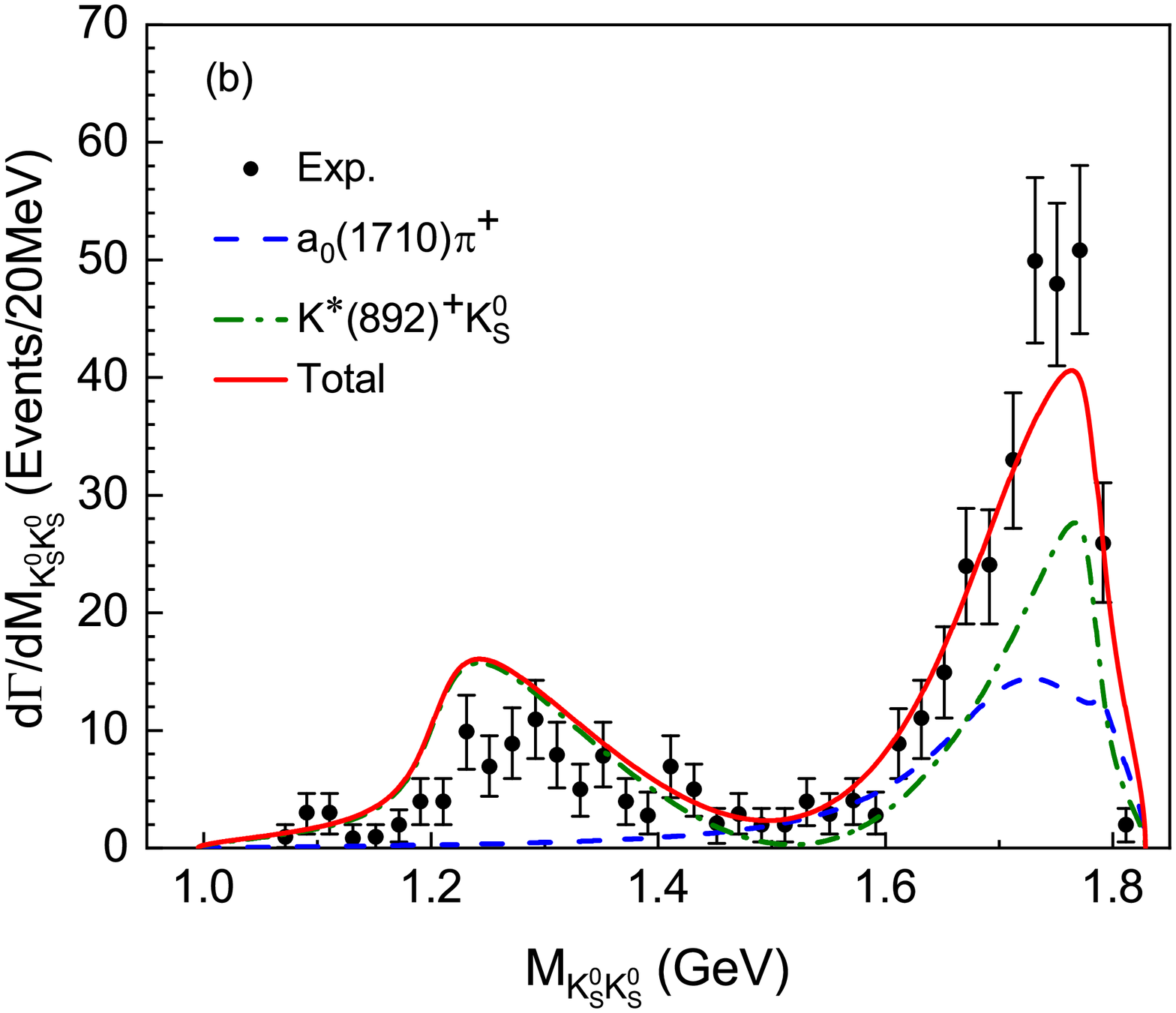}
    \end{center}
\vspace{-0.5cm} 
\caption{Invariant mass distribution of $K_S^0K_S^0$ for the $D_s^+\rightarrow\pi^+K_S^0K_S^0$ decay, compared with the experimental data taken from Fig.~2(a) of Ref.~\cite{BESIII:2021anf}. (a): results of Set I; (b): results of Set II.} \label{fig:dgdmKK}
\end{figure}

In this section, we present the numerical results for the invariant mass distribution of $K^0_SK^0_S$ and $\pi^+K^0_S$ of the $D_s^+\rightarrow\pi^+K_S^0K_S^0$ decay. To compare the theoretical invariant mass distributions with the experimental measurements, we introduce an extra global normalization factor $C$, which will be fitted to the experimental data. In Fig.~\ref{fig:dgdmKK}, we show our theoretical results for the $K^0_SK^0_S$ invariant mass distribution. The red-solid curve stands for the total contributions from the $a_0(1710)$ state and the vector $K^{*+}$ meson, while the blue-dashed and green-dot-dashed curves correspond to the contribution from only the $a_0(1710)$ and $K^{*+}$, respectively. The red-solid curve has been adjusted to the strength of the experimental data of BESIII~\cite{BESIII:2021anf} at its peak by taking $C=2.3 \times 10^7$ for both Set I and Set II. One can see that the model results  obtained with the parameters of both Set I and Set II can reproduce the experimental data reasonably well, and the $K^{*}$ plays an important role around the peak of the $a_0(1710)$ state. It is clearly seen that the shape of $a_0(1710)$ in Fig.~\ref{fig:dgdmKK} (b) is wider than that in Fig.~\ref{fig:dgdmKK} (a). One reason is that, as shown in Table~\ref{tab:azeroparameters}, the $a_0(1710)$ width of Set II is larger than the one of Set I. The other reason is that the loop function $G_{K^*\bar{K}^*}$ obtained with the cutoff regularization  of Ref.~\cite{Wang:2022pin} is  smoother than the one of Ref.~\cite{Geng:2008gx}, which can be seen in Fig.~\ref{fig:GKstarKbarstar}. 

 In Ref.~\cite{Dai:2021owu}, the vector-vector intermediate states were produced at the first step with both the external and internal $W$-emission mechanisms, and then the final-state interaction of vector-vector produces $f_0(1710)$ and $a_0(1710)$ and then they decay into $K^0_SK^0_S$ and $K^+K^-$. By adjusting the effective parameters between these production processes, the ratio of the branching fractions Br$(D^+_s \to \pi^+ K^0_SK^0_S)$ and Br$(D^+_s \to \pi^+K^+K^-)$ from the $f_0(1710)$ and $a_0(1710)$ contribution can be reproduced~\cite{Dai:2021owu}. Clearly, this work and Ref.~\cite{Dai:2021owu} share the same mechanism for the final-state interactions. As a result, both can describe the main feature of the $K^0_SK^0_S$ line shapes. In principle, both the external and internal $W$-emission mechanisms can play a role. However, a quantitative consideration of both  mechanisms inevitably introduces additional free parameters for the weak interaction (more details can be found in Ref.~\cite{Dai:2021owu}), which cannot yet be well determined. Hence, we will leave a simultaneous consideration of both mechanisms to a future study when more precise experimental data become available.

It should be noted that the contribution of the $f_0(1710)$ state is not considered in our calculation, while the data, on the other hand, contain the contributions of both states. This implies that the peaks of $f_0(1710)$ and $a_0(1710)$  overlap strongly. Otherwise, the sole contribution from $a_0(1710)$ cannot describe the experimental data. The $f_0(1710)$ and $a_0(1710)$ mixing can also be  studied in the $J/\psi$ decays~\cite{Molina:2019wjj} when more experimental data are available, just as the $a_0(980)$ and $f_0(980)$ mixing~\cite{BESIII:2018ozj,Wu:2008hx,Wu:2007jh} for the case of $K\bar{K}$ molecules. In fact, the $a_0(980)$ and $f_0(980)$ mixing was investigated in the decay $D^+_s \to \eta \pi^0 \pi^+$ in Ref.~\cite{Achasov:2017edm} with the formalism built in a earlier work of Ref.~\cite{Achasov:1979xc}, where the mixing of $a_0(980)$ and $f_0(980)$ resonances that breaks the isospin invariance due to the $K^+$ and $K^0$ meson mass difference.

In our model, the final $K^0_SK^0_S$ pair is produced from the $K^*{\bar K}^*$ interaction, and the loop function $G_{K^*\bar{K}^*}$ is very small around the  $a_0(980)$ [$f_0(980)$] pole region (see Fig.~\ref{fig:GKstarKbarstar}) and the coupling of $a_0(980)$ [$f_0(980)$] to the $K^*\bar{K}^*$ channel~\cite{Garcia-Recio:2013uva} is also small compared with the one of $a_0(1710)$ to the $K^*\bar{K}^*$ channel. Hence, there are no $a_0(980)$ and $f_0(980)$ signal in the $K^0_SK^0_S$ mass spectrum of the $D_s^+\rightarrow\pi^+K_S^0K_S^0$ decay. On the other hand, from Eq.~\eqref{eq:KstarKbarstarcombination}, we find that the two phases in the $a_0(1710)$ and $f_0(1710)$ productions have an opposite sign. If the production of $a_0(1710)$ is constructive as shown in the new BESIII data~\cite{BESIII:2021anf}, one can expect no contribution from the mechanism shown in Fig.~\ref{Fig:Feynmanquark} to produce the $f_0(1710)$ resonance in the $D^+_s \to \pi^+ K^+K^-$ decay~\cite{BESIII:2020ctr,Duan:2020vye,Wang:2021ews}. However, the $f_0(1710)$ could be produced via the external $W$ emission mechanism, and its signal is expected in the process $D^+_s \to \pi^+ K^+ K^-$. 

\begin{figure}[tbhp]
\begin{center}
\includegraphics[scale=0.32]{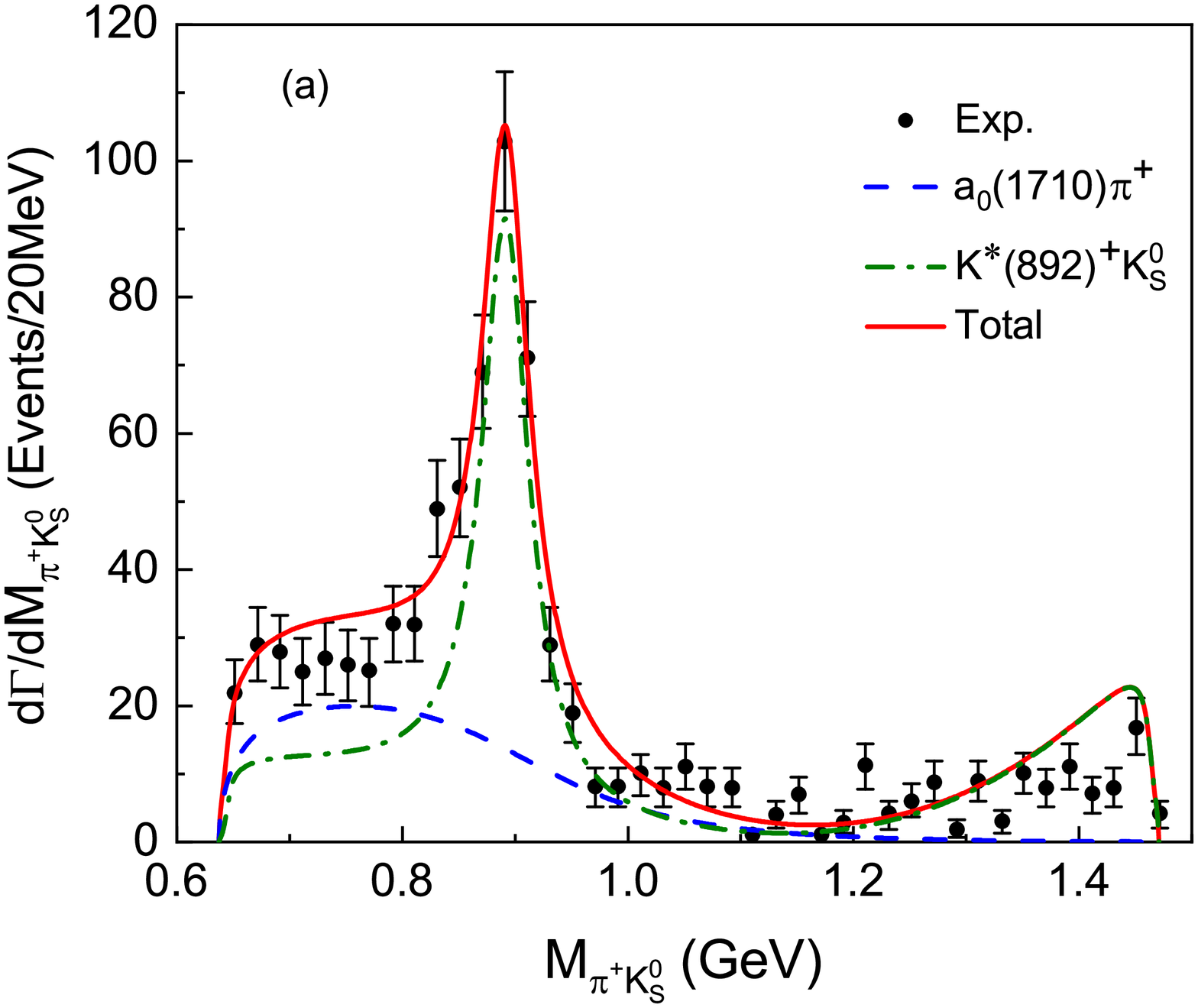}
\includegraphics[scale=0.32]{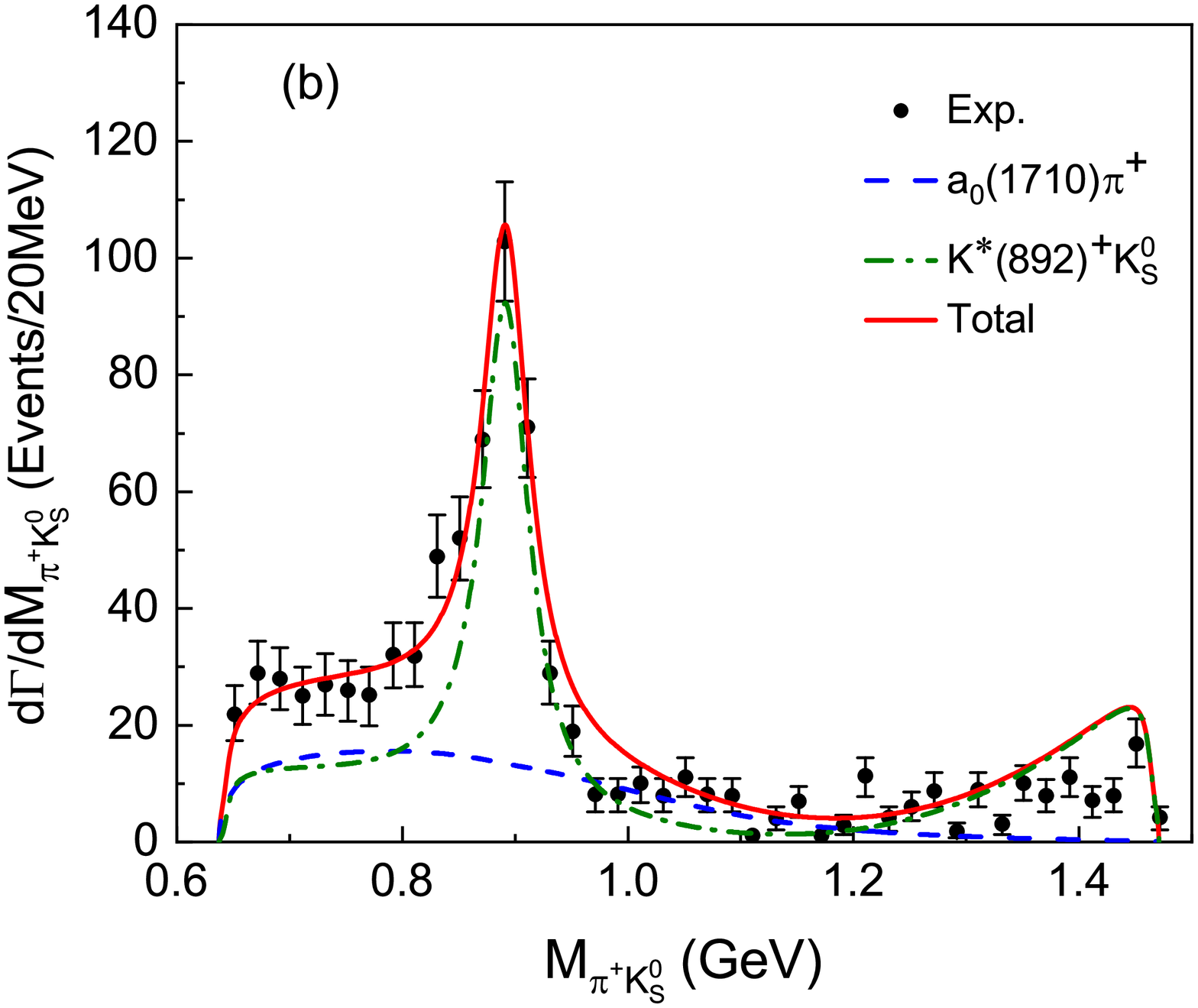}
\end{center}
\vspace{-0.5cm}
\caption{Invariant mass distribution of $\pi^+ K_S^0$ for the $D_s^+\rightarrow\pi^+K_S^0K_S^0$ decay, compared with the experimental data taken from Fig.~2(b) of Ref.~\cite{BESIII:2021anf}. (a): results of Set I; (b): results of Set II.} \label{fig:dgdmpionK}
\end{figure}

Next, we turn to the $\pi^+K^0_S$ invariant mass distributions. In Fig.~\ref{fig:dgdmpionK} we show the theoretical results for the invariant $\pi^+ K_S^0$ mass distributions of the $D_s^+\rightarrow\pi^+K_S^0K_S^0$ decay. To compare with the experimental results, we have multiplied a factor of two to ${d\Gamma}/{dM_{{\pi}K_S^0}}$, since the experimental distribution of $\pi^+K^0_S$ contains two entries of events, one for each $K^0_S$ (see more details in Ref.~\cite{BESIII:2021anf}). The peak of the $K^{*+}$ can be well described. The contribution from $a_0(1710)$ is very small to the peak, while its contribution to the threshold enhancement of the invariant $\pi^+K^0_S$ mass distribution is significant.

\begin{figure}[htbp]
    \begin{center}
        \includegraphics[scale=0.32]{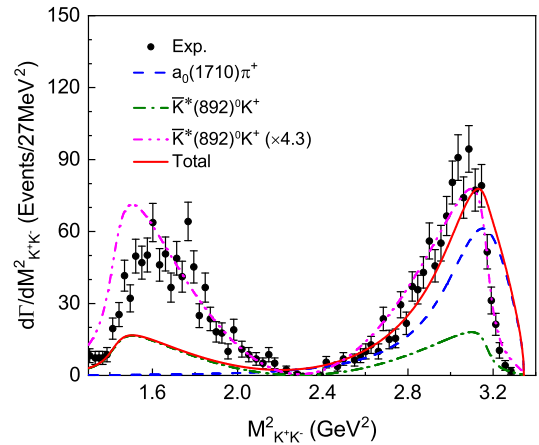}
    \end{center}
\vspace{-0.5cm} 
\caption{Invariant mass distribution of $K^+ K^-$ for the $D_s^+ \rightarrow \pi^+ K^+ K^-$ decay, compared with the experimental data taken from Fig.~7(a) of Ref.~\cite{BESIII:2020ctr}.} \label{fig:dgdmKK-charge}
\end{figure}

\begin{figure}[htbp]
\begin{center}
\includegraphics[scale=0.32]{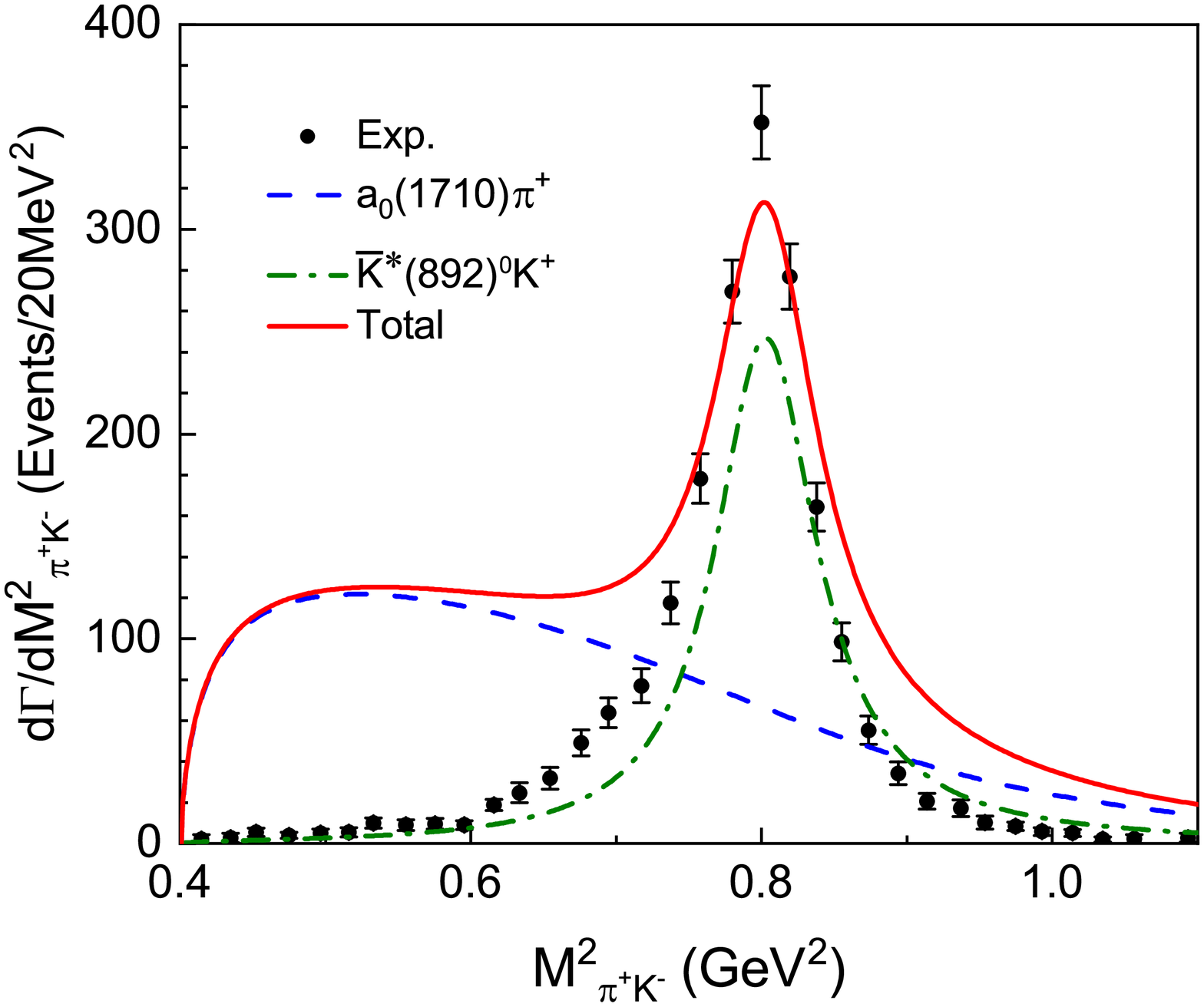}
\end{center}
\vspace{-0.5cm}
\caption{Invariant mass distribution of $\pi^+ K^-$ for the $D_s^+\rightarrow\pi^+ K^+ K^-$ decay, compared with the experimental data taken from Fig.~7(c) of Ref.~\cite{BESIII:2020ctr}.} \label{fig:dgdmpionK-charge}
\end{figure}

In addition, with the model parameters as obtained above for the $D_s^+\rightarrow\pi^+K_S^0K_S^0$ decay, we study the process of $D_s^+ \rightarrow \pi^+ K^+ K^-$, where the contributions from $a_0(1710)$ and $K^*(892)$ are taken into account by assuming that the mechanism of $D_s^+ \rightarrow \pi^+ K^+ K^-$ is the same as the one of the process of $D_s^+ \rightarrow \pi^+ K^0_S K^0_S$. The numerical results for the $K^+K^-$ and $\pi^+K^-$ invariant mass distributions are shown in Figs.~\ref{fig:dgdmKK-charge} and ~\ref{fig:dgdmpionK-charge}, respectively. The experimental data are taken from Ref.~\cite{BESIII:2020ctr}. 
If the $a_0(1710)$ plays the dominant role for the structure around $M^2_{K^+K^-}=3$~GeV$^2$, the  $\pi^+K^-$ invariant mass distribution in the low energy region can not be well described because of the reflection effect of $a_0(1710)$. 
In Fig.~\ref{fig:dgdmKK-charge}, the $K^*(892)$ contribution is scaled by a factor of 4.3, as shown by the  pink-dash-dashed curve, 
and one can see that the $K^*(892)$ contribution is already enough to reasonably describe both the $K^+K^-$ and $\pi^+K^-$ invariant mass distributions in the energy region considered. which is also consistent with the Dalitz Plot of $D_s^+ \rightarrow \pi^+ K^+ K^-$, as shown in Fig.~6 of Ref.~\cite{BESIII:2020ctr}.

As discussed above, to describe well both $D^+_s \to \pi^+ K^0_SK^0_S$ and $D^+_s \to \pi^+ K^+ K^-$ reactions, one needs to consider other mechanisms, especially the contribution from $f_0(1710)$. In this case, we will have more free parameters, and we need more constraints from both theoretical and experimental sides. In the present work, we focus on the role played by the $a_0(1710)$ in the $D^+_s \to \pi^+ K^0_S K^0_S$ decay, and it is found that the new measurements of the $D^+_s \to \pi^+ K^0_S K^0_S$ reaction can be well reproduced, and the contribution of the $K^*(892)$ around the  $a_0(1710)/f_0(1710)$ peak is important.

Finally, it is interesting to note that one can study the $a^+_0(1710)$ state in the $K^+K^0_S$ channel of the $D^+_s \to \pi^0 K^+K^0_S$ decay by including the contribution of the final-state interaction of $K^{*+}\bar{K}^{*0}$, which can be easily obtained by summing the second term of Eqs.~\eqref{eq:vp32} and \eqref{eq:pv13}. If the very small mass difference of charged and neutral $K^*$ meson is neglected, it is expected that the branching fraction of $D^+_s \to \pi^0 K^+K^0_S$ should be the same as the one of $D^+_s \to \pi^+ K^0_SK^0_S$, and hence a charged $a_0^+(1710)$ will be visible in the invariant $K^+K^0_S$ mass spectrum. Indeed, the $a_0^+(1710)$ was recently observed in the decay of $D^+_s \to \pi^0 K^+K^0_S$ by the BESIII Collaboration~\cite{BESIII:2022wkv}.

\section{Summary} \label{sec:Conclusions}

In summary, we have studied the Cabibbo-favored process of $D_s^+\rightarrow\pi^+K_S^0K_S^0$. 
By considering the decay mechanism of internal $W^+$ emission, and hadronization of the $s\bar{d}$ or $u\bar{s}$ with $q\bar{q}$ with the vacuum quantum numbers, we obtain $\pi^+K^*\bar{K}^*$ in the first step, then the transition of $K^*\bar{K}^* \to K^0_SK^0_S$  proceeds following final-state interactions of the $K^*\bar{K}^*$ pair in the chiral unitary approach where the $a_0(1710)$ state is dynamically generated. In addition, the tree diagram of $K^{*+} \to \pi^+ K^0_S$ is also taken into account.

We have calculated the  $K^0_SK^0_S$ and $\pi^+K^0_S$ invariant mass distributions, which are in good agreement with the experimental measurements of BESIII~\cite{BESIII:2021anf}. We have found that the $K^*$ plays an important role in the  $a_0(1710)$ peak region. Our study shows that the BESIII measurements support the $K^*\bar{K}^*$ molecular nature of the $a_0(1710)$ and $f_0(1710)$ states and they overlap strongly in the data.

For the reproduction of the $a_0(1710)$ peak, it is found that  the contributions from both the
tree diagram as shown in Fig.~\ref{fig:mb} and the $K^*\bar{K}^*$ final-state interaction as shown in Fig.~\ref{fig:hadronFSI} are crucial. In addition, within the proposed mechanism, it is expected that the charged $a^+_0(1710)$ signal can show up  in the $K^+K^0_S$ invariant mass distribution of the $D^+_s \to \pi^0 K^+K^0_S$ decay, which should be checked by future experimental measurements.

\section*{Acknowledgments}

We would like to thank Prof. Eulogio Oset and Prof. Lian-Rong Dai for their careful reading of the manuscript and useful discussions. This work is partly supported by the National Natural
Science Foundation of China under Grants Nos. 12075288, 11735003, 11975041,  11961141004, 11961141012, and 12192263.
This work is supported by the Natural Science Foundation of Henan under Grand No. 222300420554, the Key Research Projects of Henan Higher Education Institutions under No. 20A140027, the Project of Youth Backbone Teachers of Colleges and Universities of Henan Province (2020GGJS017), the Youth Talent Support Project of Henan (2021HYTP002), the Fundamental Research Cultivation Fund for Young Teachers of Zhengzhou University (JC202041042), and the Open Project of Guangxi Key Laboratory of Nuclear Physics and Nuclear Technology, No. NLK2021-08.


\end{document}